\documentclass[conference]{IEEEtran}
\IEEEoverridecommandlockouts
\usepackage{amsmath}
\usepackage{}
\usepackage{verbatim} 
\usepackage{pifont}
\usepackage{times,amsmath,color,amssymb,graphicx, epsfig,cite,psfrag,subfigure,algorithm}
\usepackage[noend]{algpseudocode}
\usepackage{enumerate}
\usepackage{xcolor}

\usepackage{color}
\begin{document}
	
	
	%
	\title{
		Data Volume-aware Computation Task Scheduling for Smart Grid Data Analytic Applications
 {\footnotesize 
 }
 	\thanks{This work is supported by the Natural Science Foundation of China (61931017). The corresponding author is Hongyan Li. }
 	\thanks{	
 		The source code is publicly available at https://github.com/wilixx/ICCTS.
 }
	}
        \author{\IEEEauthorblockN{Binquan Guo$^*$$\dagger$, Hongyan Li$^*$, Ye Yan$^\dagger$, Zhou Zhang$^\dagger$, and Peng~Wang$^*$}\\
        	\IEEEauthorblockA{$^*$State Key Laboratory of Integrated Service Networks, Xidian University, Xi'an, P. R. China\\
        		$\dagger$Tianjin Artificial Intelligence Innovation Center (TAIIC), Tianjin, P. R. China\\
        		Email:bqguo@stu.xidian.edu.cn, hyli@xidian.edu.cn, yanye1971@sohu.com, zt.sy1986@163.com, pengwangclz@163.com}}

	
	%
	


	\maketitle
	
	
	\begin{abstract}
	Emerging smart grid applications analyze large amounts of data collected from millions of meters and systems to facilitate distributed monitoring and real-time control tasks.
	However, current parallel data processing systems are designed for common applications, unaware of the massive volume of the collected data, causing long data transfer delay during the computation and slow response time of smart grid systems. 
	A promising direction to reduce delay is to jointly schedule computation tasks and data transfers. 
	We identify that the smart grid data analytic jobs require the intermediate data among different computation stages to be transmitted orderly to avoid network congestion. This new feature prevents current scheduling algorithms from being efficient. 
	In this work, an integrated computing and communication task scheduling scheme is proposed. The mathematical formulation of smart grid data analytic jobs scheduling  problem is given, which is unsolvable by existing optimization methods due to the strongly coupled constraints. Several techniques are combined to linearize it for adapting the Branch and Cut method. Based on the topological information in the job graph, the Topology Aware Branch and Cut method is further proposed to speed up searching for optimal solutions. Numerical results demonstrate the effectiveness of the proposed method. 
	\end{abstract}
	

	%
	\IEEEpeerreviewmaketitle

	\section{Introduction}
	
	The smart grid uses smart meters, sensors to collect data, and adopts information technologies to make smart decisions to fulfill the demand and supply of modern electrical power \cite{li2020communication}. To facilitate distributed monitoring and real-time control tasks \cite{cosovic20175g}, a huge amount of raw data are collected in real time from smart meters and sensors deployed in different geographical areas, and uploaded periodically on a hourly, daily or monthly basis (depending on the customers and purposes) to computing systems for data analysis \cite{bera2014cloud}. 
	Depending on different functions and purposes, data sources in smart grid systems may include data from phasor measurement units, power consumption patterns and data measured by the smart meters of the advanced metering infrastructure, power market pricing and bidding data, and power system equipment monitoring, control, maintenance, automation, and management data. A typical use case of smart grid data analytic applications can be found in \cite{mashima2019s}.
	In this context, the accumulated, voluminous and continuously generated data, whose volume is quite larger than that of traditional ones in common data processing systems, make real-time data analysis in smart grid systems very challenging \cite{ghorbanian2019big}. 
	According to \cite{rusitschka2010smart}, a smart grid system with 2 million customers will generate about 22 Gigabytes of data each day.
	To efficiently handle such massive data, various computing and communication optimization technologies are proposed to improve the performance, such as fog computing for energy consumption scheduling \cite{chouikhi2022energy}, software define networking in smart grid for resilient communications \cite{aydeger2016software}. Moreover, many standardization activities supported by governments and stakeholders are making continuous efforts to makes use of advanced information, control, and communications technologies in smart grids systems to save energy, reduce cost and increase reliability and transparency \cite{fan2012smart}.

	Traditionally, the typical data processing architectures like MapReduce\cite{dean2008mapreduce}, Pregel \cite{malewicz2010pregel} and Spark \cite{zaharia2012resilient}, designed for general applications, usually partition input data over a number of parallel machines, such that a data analytic job is decomposed into multiple tasks. Before generating the final results, the partial results between the adjacent stages of computation need to be exchanged through the network during the job execution. These systems developed for common purposes, focus on data partitioning and computing, and rarely optimize data transmission performance. With the rapid growth and accumulation of the data volume in smart grid, the data transfer time has become an increasingly significant bottleneck in the performance of data analytic jobs. Firstly, when tasks with precedence constraints are scheduled on different machines, the data transfer time will be increased. Secondly, when a large number of data transfers are performed at the same time, the competition with occur, leading additional delays. The increased job execution time will greatly affect the response time, which will impact the rapid decision-making ability of the smart grid systems. Therefore, optimizing data transfers is important for minimizing the job completion time, with the aim of enabling rapid decision-making for smart grid data analytic applications.

With respect to minimizing the job completion time, traditional works designed for common applications focused on 
either computation task placement (e.g., \cite{jalaparti2015network})  or network flow scheduling (e.g., \cite{wang2018survey, chen2007scheduling, soudan2009flow}). 
However, the separation between scheduling computation and communication tasks results in inefficient job processing performance especially when the data volume is huge. 
Specifically, the data transfer scheduling problem considering data volume has been well studied in \cite{chen2007scheduling, soudan2009flow}. Since their goal is to minimize network congestion, rather than reduce the overall job completion times, joint optimization of computation and data transmission has not been take into consideration.
To overcome this, some researchers recently began to break the barrier and attempted to coordinate the computation and communication tasks. The authors in \cite{jiang2016symbiosis} designed the 
Symbiosis framework to co-locate computation-intensive and network-intensive tasks, where computation tasks can utilize
the idle computing resource during the transmission of other network-intensive tasks to reduce the completion time. 	 
In the Firebird framework proposed in \cite{he2016firebird}, computation tasks
are placed based on machines' available bandwidths to avoid network contention.   
In \cite{munir2020network}, the authors proposed to place computation tasks
according to the predicted flow transfer time under given network conditions. 
In  \cite{zhao2020joint}, the authors considered  jointly	 
optimizing the reducer placement and bandwidth scheduling to minimize the coflow completion time.
Those works usually aimed to reduce the completion time
of either computation or communication tasks rather than optimize the whole job.  	 
In \cite{giroire2019network}, the authors 
considered the joint computation and communication task scheduling problem from the job perspective, and proposed
heuristic scheduling algorithms. 
In \cite{guo2022optimal}, the authors investigated the problem of joint computation and communication task scheduling with bandwidth augmentation, where mathematical optimization method was designed to solve it optimally. 
However, both of their methods are designed for general data center scenarios, which cannot be directly applied in smart grid systems.
Besides, the {mathematical model of the integrated computation and communication task scheduling problem, especially for the smart grid applications, is still missing. 
Indeed, the inherent causal relationship  between the computation task placement and data transfer condition \textcolor{black}{(i.e., data transfer may or may not be necessary depending on the placement of computation tasks)} in data analytic jobs will greatly increase the complexity of the integrated scheduling problem.

In this work, an efficient Integrated Computing and Communication Task Scheduling (ICCTS) scheme  for smart grid applications is proposed.
At first, by exploring the causal relationship, we construct the Completion Time Minimization
oriented Integrated Computation and Communication Task Scheduling
Problem (CTM-ICCTSP) to mathematically model the data analytic job scheduling problem, which is not solvable by exiting optimization tools due to a large number of complicated coupled non-linearity constraints.
Second, the general flow concept is defined to represent the internal or external data transfer
between adjacent computation tasks. Based on the general flow concept, the auxiliary virtual  channel is introduced
 to linearize  CTM-ICCSP, so that it can be directly solved by
the Branch and Cut (B\&C) method. Then, to reduce the searching space,
we utilize the topology information in the job graph model
and design a Topology Aware Branch and Cut (TABC) method to effectively speed up searching for optimal solutions.
Finally, numerical results validate
the necessity of optimizing the data transfers
as well as the effectiveness of the proposed ICCTS method.
	
	\section{System Model}
	
	In this work, we model the smart grid data analytic jobs as periodic jobs, which are executed on a hourly or daily or monthly basis depending on their purposes, and their detailed knowledges can be profiled from historical logs. Each job is represented by a Dual Weight Directed Acyclic Graph (DWDAG)
	$ G=\{\mathcal{V},\mathcal{E},P,Q,R\}$.
	$\mathcal{V}$ is the computation task set, and
	$v_j  \in \mathcal{V}$ denotes      
	the $j$-th computation task. 
	The execution of task $v_j$ lasts for $p_j$ time slots, and $P = \{ {p_j}|1 \le j \le \left| \mathcal{V} \right|\} $ is the execution time set for computation tasks.      
	$\mathcal{E}$ is the dual-weighted edge set, and edge $e_{\emph{uv}}\in\mathcal{E}$
	represents the dependency between the computation task $u$ and $v$, which means the execution of task $v$ requires the data from $u$.
	The two weights of a edge separately represent the internal and external data transfer time corresponding to the  different placement of  precedence-constrained computation tasks.
	If the pair of precedence-constrained computation tasks, denoted by $u$ and $v$,
	are placed in the same machine, the intermediate data on edge
	$e_{\emph{uv}}$
	are transmitted internally, and the network flow will not occur;
	otherwise flow $f_{\emph{uv}}$ on edge
	$e_{\emph{uv}}$ 
	will be transmitted externally through the communication channel between  machines accommodating  $u$ and $v$. The corresponding internal  and external transfer times are separately denoted by
	$r_{e_{\emph{uv} }} \in R$ and $q_{e_{\emph{uv} }} \in Q$.

     Fig. 1 illustrates a DWDAG example with six computation tasks and eight
     possible network flows.  For a job, a set of available machines are reserved to process its data, denoted by  $ \mathcal{M}=\{\alpha_i | 1 \le i \le M \}$, while the  network resource shared among these machines is modeled by a set of individual communication channels with the same bandwidth, denoted by $ \mathcal{N}=\{\beta_k | 1 \le k \le N \} $.}
 The time axis is cut into multiple identical slots, and the slot index set is denoted as 
$ \mathcal{T}=\{\tau | 1 \le \tau \le T_{\emph{max}} \} $.
   We assume that  each computation task's 
   processing time and each data's transfer time are predetermined,
   and computation tasks and network flows are
   scheduled non-preemptively, i.e.,
   once started,
   their processing or transmission cannot be interrupted.

		\begin{figure}
		\centering
		\includegraphics[width=80mm]{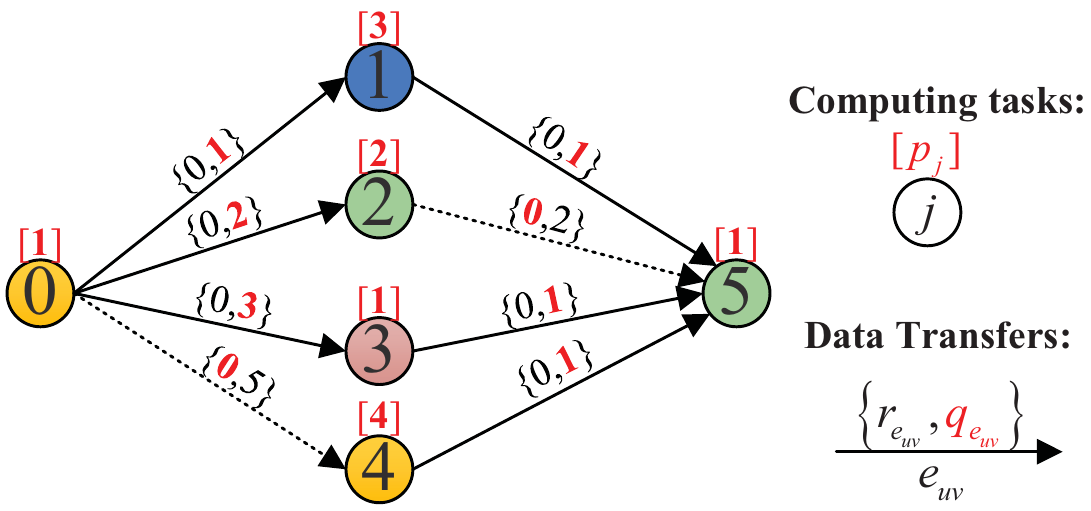}
		\caption{A smart grid data analytic job example  with six computation tasks and eight possible data transfers (i.e., network flows).}
	\end{figure}

 \section{Integrated Computation and Communication Task Scheduling Problem}
	 	In this section, the smart grid data analytic job scheduling problem is formulated as a Completion Time Minimization oriented Integrated Computation and Communication Task Scheduling Problem (CTM-ICCTSP).
The computation task set and possible network flow set are denoted by $\hat{\mathcal{J}}$ and
$\mathcal{F}$, respectively.
For single job scheduling, we have  $\hat{\mathcal{J}} = \mathcal{V}$ and $\mathcal{F} = \{f_{\emph{uv} }|e_{\emph{uv} } \in  \mathcal{E}\}$. 
The external flow
set $\hat{\mathcal{F} }\subseteq \mathcal{F}$ only contains the
  flows transferred via the  external physical
links between machines.

The binary computation task placement decision variable is denoted by
$\mathcal{X}_{ji\tau}$.
If task $v_j$ begins to be executed at time $ {\tau }$ in machine $\alpha_i$, ${\mathcal{X}_{ji\tau} = 1}$; 
otherwise ${\mathcal{X}_{ji\tau} = 0}$.
Since each task can be executed only once, we have
	\begin{equation}\label{c_taskCompletion}
	\sum_{1 \le i \le M} \sum_{\tau \in \mathcal{T}} {\mathcal{X}_{ji\tau}}  = 1, \forall v_j \in \hat{\mathcal{J}}.
	\end{equation}
	Similarly, the binary network flow scheduling decision variable
${\mathcal{Y}_{fk\tau} = 1}$ means  flow $ {f}$
begins to transmit at time $ {\tau }$ in communication channel $\beta_{k}$, and thus we have
	\begin{equation}\label{c_necessaryFlowCompletion}
	\sum_{1 \le k \le N} \sum_{\tau \in \mathcal{T}} {\mathcal{Y}_{fk\tau}} = 1, \forall f \in \hat{\mathcal{F} }.
	\end{equation}

The start time of computation task $v_j$ and  flow $f$,
separately denoted by $ s_j^{\mathcal{M}} $ and $ s_f^{\mathcal{N}}$, are calculated as follows.
	\begin{equation*}
	s_j^{\mathcal{M}} \overset{\Delta}{=} \sum_{\tau \in \mathcal{T}}  \sum_{1 \le i \le M} \tau \mathcal{X}_{ji\tau}, \   \  s_f^{\mathcal{N}} \overset{\Delta}{=} \sum_{\tau \in \mathcal{T}} \sum_{1 \le k \le N}  \tau \mathcal{Y}_{fk\tau}.
	\end{equation*}

 The optimization objective is to
	minimize the job completion time, i.e., the maximum completion time among all computation and communication tasks:	
\begin{equation*}
\mathop {\min }\limits_{{{\cal X}_{ji\tau }}, {{\cal Y}_{fk\tau }}} \; C_{\emph{max}}=\max \left\{ {s_j^{\cal M} + {p_j}} \right\}.
	\end{equation*}

The number of the machine to place $v_j$ is denoted by
\begin{equation*}
m_j \overset{\Delta}{=} \sum_{1 \le i \le M} \sum_{\tau \in \mathcal{T}} i \cdot \mathcal{X}_{ji\tau},
\end{equation*}
and the number of communication channel to send $f$ is denoted by
\begin{equation*}
n_f \overset{\Delta}{=} \sum_{1 \le k \le N} \sum_{\tau \in \mathcal{T}} k \cdot \mathcal{Y}_{fk\tau}.
\end{equation*}

In the  smart grid data analytic job scheduling problem, the following inherent constraints must be
carefully taken into consideration.

	\textit{Computing resource constraints:}
   The computing resource constraints include (\ref{c_taskStartTimeNonNegative}) and (\ref{c_taskDisjunctiveConstraints}).
   Constraint (\ref{c_taskDisjunctiveConstraints}) means that the execution times of
   computation tasks in the same machine will not  overlap, which is a typical  disjunctive
   constraint.
   	\begin{equation}\label{c_taskStartTimeNonNegative}
	s_j^{\mathcal{M}} \geq 0, \forall v_j \in \hat{\mathcal{J}}  \\
	\end{equation}
	\begin{equation}\label{c_taskDisjunctiveConstraints}
	s_j^{\mathcal{M}} +  p_{j} \leq  s_{j'}^{\mathcal{M}} \text{ or }  s_{j'}^{\mathcal{M}} +  p_{j'} \leq  s_{j}^{\mathcal{M}},\forall   v_j \neq v_{j'}, m_j = m_{j'} \\
	\end{equation}

	\textit{Causality constraints:} The occurrence of flow $f_{\emph{uv} }$ depends on the placement of
 its precedence-constrained tasks $u$ and $v$.
	Thus the external flow 
	set $\hat{\mathcal{F}}$ can be rewritten in
the causal relationship based formulation as

\begin{equation}\label{c_causalRelationship}
\hat{\mathcal{F}} = \{f_{\emph{uv} }|f_{\emph{uv} } \in  \mathcal{F}, m_u \neq m_v\}.
\end{equation}

	\textit{Precedence constraints:}
     The precedence relationship exists between two  adjacent
     upstream and downstream computation tasks.
     A task starts to be executed after all of its precedent tasks and upstream necessary flows end.
     Depending on where precedence-constrained tasks are placed
     in the same machine,  two cases should be considered.

	\textbf{Case 1:}
	If task $u$ and $v$ are placed on the same machine, the data between $u$ and $v$ is
transferred within the machine, and  the  start  time
of the downstream task $v$  should be after   $u$'s end time adding 
the internal transfer time, denoted by
	\begin{equation}\label{c_precedenceConstraintCase1}
	s_u^{\mathcal{M}}  + p_u + r_{e_{\emph{uv}}} \leq  s_v^{\mathcal{M}}  , \forall  f_{\emph{uv}} \in \mathcal{F}-\hat{\mathcal{F}}.
	\end{equation}

	\textbf{Case 2:}
If task $u$ and $v$ are placed on different machines, the flow between $u$ and $v$ is
transferred between the machines. Thus the start time of $f_{\emph{uv} }$ should be after $u$'s end time, while $v$'s start  time should be after the end time of $f_{\emph{uv} }$, denoted by
(\ref{c_precedenceConstraintCase21}) and (\ref{c_precedenceConstraintCase22}).
	\begin{equation}\label{c_precedenceConstraintCase21}
	s_{u}^{\mathcal{M}}  + p_u \leq  s_{f_{\emph{uv} }}^{\mathcal{N}}, \forall   f_{\emph{uv} } \in \hat{\mathcal{F}}
	\end{equation}
	\begin{equation}\label{c_precedenceConstraintCase22}
	s_{f_{\emph{uv} }}^{\mathcal{N}}  + q_{e_{uv}} \leq  s_{v}^{\mathcal{M}}, \forall  f_{\emph{uv} } \in \hat{\mathcal{F}}
	\end{equation}

	\textit{Communication resource constraints: }
	The Communication resource constraints include (\ref{c_flowStartTimeNonNegative}) and (\ref{c_flowDisjunctiveConstraints}).
	Constraint (\ref{c_flowDisjunctiveConstraints}) means that the  flow transmissions in the same network channel will not overlap, which is also a typical  disjunctive
	constraint.
	\begin{equation}\label{c_flowStartTimeNonNegative}
	s_{f}^{\mathcal{N}} > 0, \forall f \in \hat{\mathcal{F}}
	\end{equation}
	\begin{small}
\begin{equation}\label{c_flowDisjunctiveConstraints}
s_{f}^{\mathcal{N}} +  q_{f'} \leq s_{f'}^{\mathcal{N}} \text{ or } s_{f'}^{\mathcal{N}} +  q_{f'} \leq s_{f}^{\mathcal{N}}, \forall f,f' \in \hat{\mathcal{F}}, f\neq f', n_f = n_{f'}
\end{equation}
	\end{small}

	Finally, the resulting CTM-ICCTSP can be expressed as follows.
	\begin{align*}
	& \mathbf{P1: }\text{ min } C_{\emph{max}}\\
	& \text{s.t. }  (\ref{c_taskCompletion})-(\ref{c_flowDisjunctiveConstraints})
	\end{align*}

	\section{Linearized Reformulation and Topology Aware Branch and Cut Method}

	\subsection{\textcolor{black}{General flow concept and auxiliary virtual  channel}}
  Due to constraints (\ref{c_taskDisjunctiveConstraints}),  (\ref{c_causalRelationship}), (\ref{c_precedenceConstraintCase1}), (\ref{c_precedenceConstraintCase21}), (\ref{c_precedenceConstraintCase22}) and  (\ref{c_flowDisjunctiveConstraints}), CTM-ICCTSP is a non-linear problem. Constrains
    (\ref{c_taskDisjunctiveConstraints}) and (\ref{c_flowDisjunctiveConstraints}) can be transferred into linear constraint formulations by  the Big-M and Convex Hull reformulation methods in disjunctive programming.
However, constraint (\ref{c_causalRelationship}) is not so easy to handle.
	To eliminate the volatility of
      $\hat{\mathcal{F}}$ in  (\ref{c_causalRelationship}), we define the general flow concept.
     A general flow may be an internal or external flow transferred between
     two precedence-constrained tasks.
     Only the external flows compete for network resource. To construct a unified
     flow scheduling framework compatible with the two  types of flows,
     we introduce the auxiliary virtual channel, which is contention-free for
     all internal flow transfers. Thus the external flows are transferred via physical network links
     while the internal flows are handled by the auxiliary virtual channel.
     By introducing this auxiliary virtual channel, 
     the uncertainty of external flows can be eliminated.
     The auxiliary virtual channel is denoted by $ \hat{k} $,
     and thus the communication resource set is $ {\mathcal{N} \cup \hat{k}} $.

	A  general flow $f \in \mathcal{F}$ must be placed on either real communication channels or the auxiliary  virtual channel, which separately corresponds to the external or internal transfer.
Since each general flow must be transferred, we have
	\begin{equation}\label{c_generalFlowCompletion}
	\sum_{\tau \in \mathcal{T}} \sum_{k\in \mathcal{N} \cup \hat{k}}^{n} {\mathcal{Y}_{fk\tau}} = 1,\forall f \in \mathcal{F}.
	\end{equation}

	\subsection{Linearization of computation and communication disjunctive constraints
 }
	By introducing the general flow and auxiliary  virtual   channel,
the scheduling entity $\hat{\mathcal{F}}$ in \textbf{P1} can be replaced by
the general flow set $\mathcal{F}$, which
eliminates the uncertainty of the
the scheduling entity.
Therefore, we can continue to adopt the
reformulation methods in disjunctive programming \cite{balas1998disjunctive}
to transform \textbf{P1} into an equivalent Integer Linear Programming (ILP) problem.
To linearize constraint (4), we introduce two types of auxiliary variables to
describe the placement of computation tasks.
Binary variables
${\psi _{jj'i}} \in \{ 0,1\}$ indicate whether two computation tasks
are placed in the same machine, where
$v_j,v_{j'} \in \widehat {\cal J}, v_j \ne v_{j'},1 \le i \le M$.
If computation tasks $v_j$ and $v_{j'}$ are both placed on the $i$-th machine,
 $ {\psi_{jj'i}} = 1 $; otherwise $ {\psi_{jj'i}} = 0$.
To construct ${\psi _{jj'i}}$, we have
	\begin{equation}\label{c_taskPlacementIndicator}
	\begin{split}
	0 & \leq \sum_{\tau \in \mathcal{T}} {\mathcal{X}_{ji\tau}} + \sum_{\tau \in \mathcal{T}} {\mathcal{X}_{j'i\tau}} - 2 \cdot \psi_{jj'i} \leq 1,  \\
	& \forall v_j,v_{j'} \in \hat{\mathcal{J}}, v_j\neq v_{j'},  1 \le i \le M.
	\end{split}
	\end{equation}

 Binary precedence indicator  variables $\sigma_{jj'} \in \{0,1\}$ represent the precedence relationship
 between two computation tasks.
 If task $v_j$ starts no later than task $v_{j'}$, $\sigma_{jj'} = 1$; otherwise $\sigma_{jj'} =0$.
These constraints can be guaranteed by
	\begin{equation}\label{c_taskPrecedenceIndicator}
	\begin{split}
	& s_{j'}^{\mathcal{M}} - s_{j}^{\mathcal{M}} \leq T_{\emph{max}} \cdot
	\sigma_{jj'} - \epsilon(1-\sigma_{jj'}),\\
	&\forall v_j,v_{j'} \in \hat{\mathcal{J}}, v_j\neq v_{j'}.
	\end{split}
	\end{equation}
	where $\epsilon \in (0,1)$ is a small enough constant commonly used in the logical formulation of integer programming.
    With ${\psi _{jj'i}}$ and $\sigma_{jj'}$, the disjunctive constraint (4) can be linearized as
    \begin{equation}\label{c_computingDisjunctiveConstraints}
    \begin{split}
	& s_{j}^{\mathcal{M}} + p_{j} - s_{j'}^{\mathcal{M}} \leq  T_{\emph{max}}\cdot  (2- \sigma_{jj'}- \sum_{1 \leq i \leq M} \psi_{jj'i}),
	\\ &\forall v_j,v_{j'} \in \hat{\mathcal{J}}, v_j\neq v_{j'}.
    \end{split}
	\end{equation}
   Constraint (\ref{c_computingDisjunctiveConstraints}) guarantees that task
	$v_{j'}$ must start after the completion of task $v_j$ when they
    are placed in the same machine, i.e.,
    $\sigma_{jj'} = 1$ and
	$ \sum_{1 \leq i \leq M} \psi_{jj'i}=1 $   hold simultaneously.

	Similarly, to linearize constraint (10),  auxiliary binary variables
$\chi_{\emph{ff}'\emph{k}}$ and $\phi_{\emph{ff}'}$ are introduced.
$\chi_{\emph{ff}'\emph{k}}$ indicates whether two flows are both placed in the $k$-th communication channel, while precedence indicator variable $\phi_{\emph{ff}'}$ represents the precedence relationship between two flows.
The constraints of $\chi_{\emph{ff'k}}$ and $\phi_{\emph{ff'}}$ as well
 as the reformulation of constraint (10) are shown as follows.
	\begin{equation}  
	\begin{split}
	& 0  \leq \sum_{\tau \in \mathcal{T}} {\mathcal{Y}_{fk\tau}} + \sum_{\tau \in \mathcal{T}} {\mathcal{Y}_{f'k\tau}} - 2 \cdot \chi_{\emph{ff}'\emph{k}}  \leq 1,\\
	& \forall f,{f'} \in \mathcal{F}, f \neq {f'}
	\end{split}
	\end{equation}

	\begin{equation}
	\begin{split}
	&  s_{f}^{\mathcal{N}} - s_{f'}^{\mathcal{N}}   \leq T_{\emph{max}} \cdot
	\phi_{\emph{ff}'} - \epsilon(1-\phi_{\emph{ff}'}), \\
	&   \forall f,{f'} \in \mathcal{F},f \neq {f'}
	\end{split}
	\end{equation}

	\begin{equation}
	\begin{split}
	&  s_{f}^{\mathcal{N}} + q_{f}  - s_{f'}^{\mathcal{N}}   \leq T_{\emph{max}} \cdot
	(2-\phi_{\emph{ff}} -\sum_{1 \leq k \leq N} \chi_{\emph{ff}'\emph{k}}), \\
	&   \forall f,{f'} \in \mathcal{F},f \neq {f'}
	\end{split}
	\end{equation}
	
	\subsection{Linearization of causality and precedence constraints between computation and communication tasks}
	
After the separate linearizations of computation and communication disjunctive
constraints, we continue to reformulate the causality and precedence constraints
(\ref{c_causalRelationship}), (\ref{c_precedenceConstraintCase1}), (\ref{c_precedenceConstraintCase21}) and (\ref{c_precedenceConstraintCase22}) to integrate the scheduling of computation and communication tasks.

With the introduced auxiliary variable ${\psi _{jj'i}}$, the
causality relationship between computation task placement and network flow occurrence
in constraint (6) can be rewritten by
	\begin{equation}
	\sum_{1 \leq i \leq M} \psi_{uvi} = \sum_{\tau \in \mathcal{T}} \mathcal{Y}_{f_{\emph{uv} },\hat{k},\tau}  , \forall {f_{\emph{uv}}} \in \mathcal{F}.
	\end{equation}

With the general flow concept, the two cases in the previous precedence constraints
can be represented in a unified form. 	
	For each pair of precedence-constrained tasks $u$ and $v$ connected by general flow $f_{\emph{uv}}$,
the start time of  $f_{\emph{uv}}$ must be after the end time of $u$, i.e.,
	\begin{equation}
	s_{u}^{\mathcal{M}}  + p_{u}  \leq
	s_{f_{\emph{uv}}}^{\mathcal{N} \cup \hat{k}},  \forall {f_{\emph{uv}}} \in \mathcal{F}.
	\end{equation}

	If flow ${f_{\emph{uv} }}$ is transferred internally,
	$\sum_{\tau \in \mathcal{T}} \mathcal{Y}_{f_{\emph{uv} },\hat{k},\tau} = 1$;
 otherwise,  the answer is zero. No matter ${f_{\emph{uv} }}$ is transferred externally or internally, the start time
 of task $v$ must be after the end time of flow ${f_{\emph{uv} }}$, denoted by
     \begin{equation}
     \begin{split}
	& s_{f_{\emph{uv}}}^{{\mathcal{N}} \cup \hat{k}}+  r_{u}\sum_{\tau \in \mathcal{T}} \mathcal{Y}_{f_{\emph{uv} },\hat{k},\tau} +
q_{f_{\emph{uv}}} (1- \sum_{\tau \in \mathcal{T}} \mathcal{Y}_{f_{\emph{uv} },\hat{k},\tau})  \leq
s_{v}^{\mathcal{M}}.
     \end{split}
	\end{equation}

	Finally,  CTM-ICCTSP can be linearized as \textbf{P2}.
	\begin{equation*}
	\begin{split}
	\mathbf{P2: }
	& \text{min } C_{\emph{max}} \\
	s.t. \ 	& (1),(3), (11)-(20),
	\\ & C_{\emph{max}} \geq  s_{j}^{\mathcal{M}} + p_j , \forall v_j \in \hat{\mathcal{J}},
	\\& \textcolor{black}{C_{\emph{max}} \geq  s_{f}^{\mathcal{N}\cup \hat{k}} + q_{e_\emph{f} }(1-\sum_{\tau \in \mathcal{T}} \mathcal{Y}_{f_{\emph{uv} },\hat{k},\tau}) , \forall f \in \mathcal{F}   },
	\\& \textcolor{black}{C_{\emph{max}} \geq  s_{f}^{\mathcal{N} \cup \hat{k}} + r_{e_\emph{f} } \sum_{\tau \in \mathcal{T}} \mathcal{Y}_{f_{\emph{uv} },\hat{k},\tau}, \forall f \in \mathcal{F}}.
	\end{split}
	\end{equation*}

	\subsection{Topology-aware Branch and Cut (TABC) Algorithm}

	Problem \textbf{P2} is an ILP problem and can be solved by the classic B\&C algorithm. Since
the searching iterations of  B\&C may vary  dramatically
	in different problem instances, 
	its running time
	may be unacceptable in some cases.	
To avoid this situation, we  take the topological relationship
among computation tasks and network flows in the DWDAG into consideration and propose an efficient Topology Aware Branch and Cut (TABC)  Algorithm based on
the following strategies.

	\subsubsection{\textcolor{black}{\textit{Branch with the chain precedence constraints}}}
	The branching process can be carried out from enumerating  the precedence indicator matrices,
${\bf{\Theta}}  = ({\sigma_{\emph{jj}'}})$ or ${\bf{\Phi }} = ({\phi _{\emph{ff}'}}{\rm{)}}$.
 Based on the precedence relationships among computation and communication tasks, some searching branches can be
 directly pruned, and thus the searching space can be greatly reduced.
 The chain precedence constraints include the  precedence constraints along successive
 computation task and flow chains.
For example, in computation tasks if $\sigma_{\emph{jj}'}=\sigma_{\emph{jj}''}=1$, then $\sigma_{\emph{jj}''}=1$.
	For network flows, if $\phi_{\emph{ff}'}=\phi_{\emph{f}'\emph{f}''}=1$, then $\phi_{\emph{ff}''}=1$. 
In general,{if  $f_{\emph{uv}} \in \mathcal{F}$, then $\sigma_{\emph{uv}}=1$; if $f_{\emph{uv}},f_{\emph{vv'}} \in \mathcal{F}$, then $\phi_{f_{\emph{uv}}f_{\emph{vv'}}}=1$.

	\subsubsection{Update task interval constraints}
	In each job's DWDAG, 
	the earliest and latest start time 
	of each computation task or network flow can be inferred according to both 
	the job's  incumbent upper and lower bounds and the processing and transfer times of the other computation tasks and flows along its longest branch. A graph theory-based method to obtain the longest branch of a directed acyclic graph can be found in \cite{guo2022optimal}, which is applicable to this work and very easy to implement.
	Therefore, the searching tree can be further pruned according to the reduced interval constraint for each task or flow.

	\subsubsection{Utilize the symmetry of solution space}
	One symmetry feature is from the homogeneity of physical machines and communication channels. 
	If we switch the indexes of two scheduled machines or communication channels in a feasible solution, the result is an equivalent solution. Setting each task's machine affinity value  in advance can eliminate a large number of symmetric solutions. 	 
	The other symmetry feature comes from the symmetry of nodes or edges in the job graph, thus different priorities are added to the equivalent computation tasks or network flows to reduce redundant searching.

	 \begin{algorithm}
	    	\caption{Topology Aware Branch and Cut Algorithm}
	    	\label{array-sum}
	    	\hspace*{0.02in} {\bf Input:}
	    	Job $G=\{\mathcal{V},\mathcal{E},P,Q,R\}$,
	    	resource set $\{\mathcal{M}\}$ and $\{\mathcal{N}\}$.\\
	    	\hspace*{0.02in} {\bf Output:}
	    	Optimal solution $S^*$.
	    	\begin{algorithmic}[1]
	    		\item {\bf Initialization:} 
	    		\item  Calculate a solution $S$ using heuristics.
	    		\item Set $\emph{LB}=\frac{sum(P)}{|M|}$, $\emph{UB}=min(C_{\emph{max}}(S),sum(P))$.
	    		\item Set the initial precedence matrices with chain precedence constraints in $G$, the initial interval constraints, and the
	    		affinities and priorities  	of   tasks/flows.
	    	    \item {\bf Repeat}
	    		\item  Solve  $\mathbf{P2}$ using B\&C for an incumbent solution.
	    		\item If new $\emph{LB}$ or $\emph{UB}$ is obtained,  recalculate and update interval constraints to the current active node in $\mathbf{P2}$.
	    		\item  {\bf Until} Optimal solution $S^*$ found.
	    		\item  {\bf return} $S^*$.
	    	\end{algorithmic}
	    \end{algorithm}

	The TABC procedures are shown in Algorithm 1.   
	The  three pruning strategies are all adopted to efficiently reduce the 	iterations.    
 Due to the inherent unreliability and instability properties of B\&C  
  in solving ILP problems, the performance
 of the proposed TABC may also vary in different cases. 
 Proper termination conditions can be set to
 avoid too long running time. 
 Though TABC may terminate  before producing a global  optimal solution, the incumbent solution can still be better than most heuristics.  In addition, since the variables in \textbf{P2} are binary, TABC can be efficiently implemented by splitting \textbf{P2} into multiple parallelly executed  sub-problems. Thus the time complexity can be greatly reduced.
 
	\section{Simulation Results}

	To evaluate the performance of the proposed scheme, we conduct simulations over the synthetic smart grid data analytic jobs. 
	As in \cite{guo2022optimal}, the computation task processing times and the data transfer times are randomly and uniformly chosen from [1, 100] and [1, 50], respectively, which is to mimic the different data volumes by normalizing the maximum time to 100.
	The larger the transfer time corresponds to the larger the data volume transfered among the adjacent tasks.

	In Fig. 2, we  compare the average normalized makespans (i.e., job completion times) of six scheduling schemes with different machine numbers and one network channel. The Random Scheduling scheme places the computation tasks randomly, 
	 while the List Scheduling scheme is from \cite{rayward1987uet}. Both of them only considered the placement of computation tasks. 
	 The Partition Scheduling, Generalized List (G-List) Scheduling and G-List-Master Scheduling schemes are from \cite{giroire2019network}.
	 For each scheduling scheme, we generated 3000 job cases each with ten computation tasks, calculated the normalized makespans of these  jobs, and averaged them. The normalized makespan is the ratio of the makespan obtained by one scheduling scheme and the upper bound makespan when the job's computation tasks  were all placed on a single machine. For the Random Scheduling and the List Scheduling schemes, their average normalized makespans increase with the machine numbers due to the ignorance of data transfer optimization. 
	 For the other four scheduling schemes, their average normalized makespans decrease with the increase of machine numbers since the computation and communication tasks are jointly scheduled.  
	 The proposed Integrated Computing and Communication Task Scheduling (ICCTS) scheme obtains the lowest average normalized makespans among all the six scheduling schemes. 	 
	 Since the linearization reformulation keeps the optimality of the original CTM-ICCTSP problem, ICCTS   
	 can obtain the optimal solution and act as an important benchmark 
	 for heuristic schemes. It can be observed that when the machines (computing resources) are sufficient, ICCTS can averagely reduce the job completion time by up to $5\%$.

	\begin{figure}[h!]
	\centering
	\includegraphics[width=75mm]{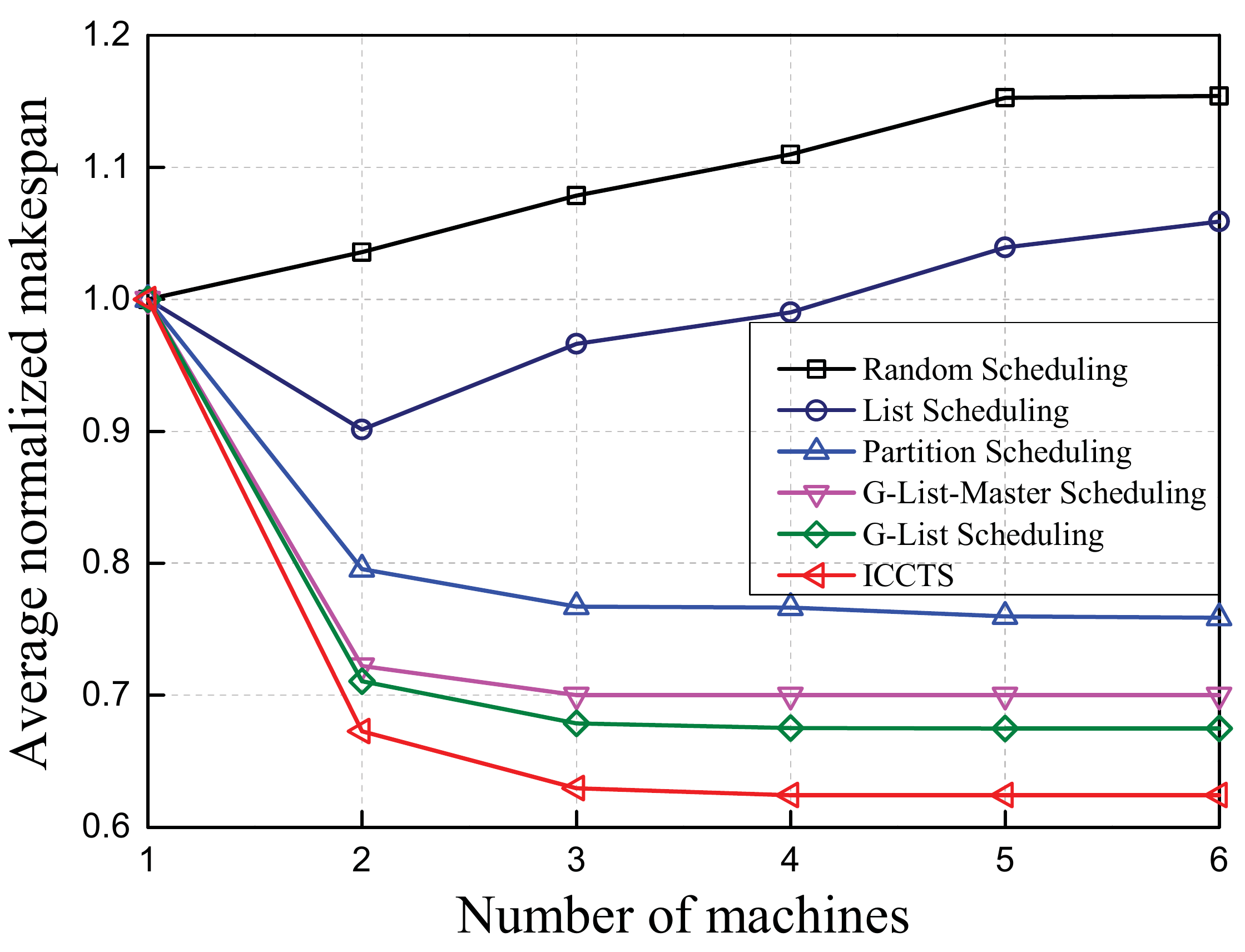}\\
	\caption{Average normalized makespans of different scheduling schemes versus the number of machines.}
\end{figure}

In Fig. 3, we compare the efficiency  of  B\&C and TABC in the  ICCTS scheme, and the simplex iteration number is adopted as the algorithm efficiency metric. Though their average simplex iterations both increases exponentially with the computation task number, TABC can significantly reduce the  iterations due to the pruning rules from DWDAG.

\begin{figure}[h!]
	\centering
	\includegraphics[width=75mm]{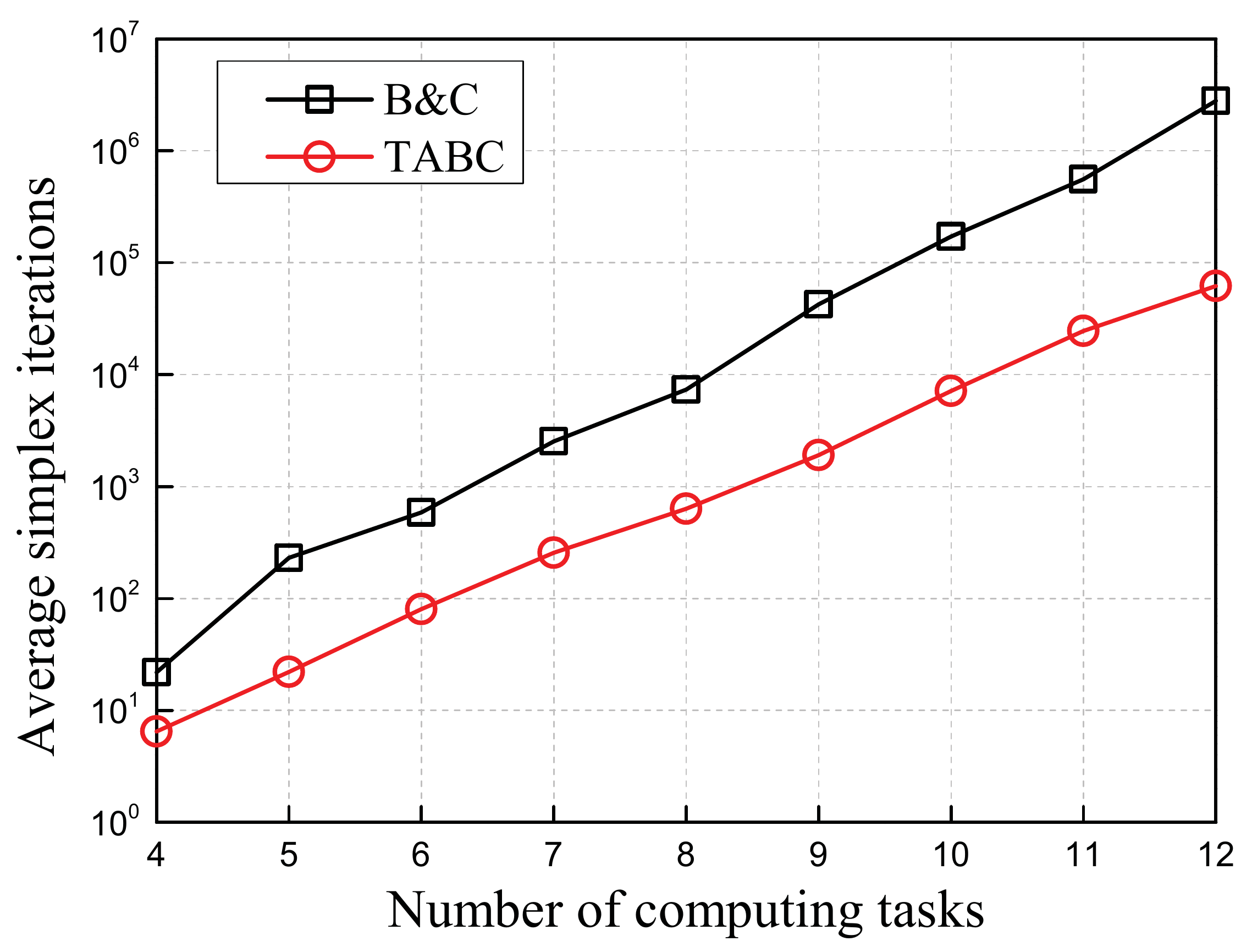}\\
	\caption{ Average simplex iterations of B\&C and
		TABC versus the number of computing tasks.}
\end{figure}

	\section{Conclusion}
	In this work, an integrated computation and communication task scheduling scheme for smart grid data analytic applications is proposed. The mathematical formulation and the corresponding constraint linearization of the job scheduling problem were introduced, and  an efficient Topology Aware Branch and Cut method was designed to improve the searching speed for the optimal solutions. Numerical results confirmed the necessity of considering data volume and validity of the proposed integrated scheduling scheme.

	
	
	%

	


	\ifCLASSOPTIONcaptionsoff
	\newpage
	\fi

	
	
	%
	
	\bibliographystyle{IEEEtran}
	\bibliography{reference}

\end{document}